\documentclass[11pt,fleqn,twoside]{article}
\usepackage{amsfonts,latexsym}
\makeatletter
\newcommand{\prava}{\footnotesize\it
\begin{flushright}
\begin{minipage}{18cm}
Copyright \copyright 1998 by W.W. Zachary and V.M. Shtelen
\end{minipage}
\end{flushright}}

\newcommand{\name}[1]{\begin{flushleft}
                       \LARGE \bf #1
                       \end{flushleft}\vspace{-3mm}}

\newcommand{\Author}[1]{\begin{flushleft}
                       \it #1 \end{flushleft}}

\newcommand{\Adress}[1]{\begin{flushleft}
                       \it #1 \end{flushleft}}

\newcommand{\Date}[1]{\begin{flushleft}
                      \small  \it #1 \end{flushleft}}

\newcommand{\ehkol}{Author \ name}
\newcommand{\ohkol}{Article \ name}
\renewcommand{\@evenhead}{
\hspace*{-3pt}\raisebox{-15pt}[\headheight][0pt]{\vbox{\hbox to \textwidth
{\thepage \hfil \ehkol}\vskip4pt \hrule}}}
\renewcommand{\@oddhead}{
\hspace*{-3pt}\raisebox{-15pt}[\headheight][0pt]{\vbox{\hbox to \textwidth
{\ohkol \hfil \thepage}\vskip4pt\hrule}}}
\renewcommand{\@evenfoot}{}
\renewcommand{\@oddfoot}{}

\newcommand{\be}{\begin{equation}}
\newcommand{\ee}{\end{equation}}
\newcommand{\ba}{\hspace*{-5pt}\begin{array}}
\newcommand{\ea}{\end{array}}

\newcommand{\ds}{\displaystyle}
\makeatother

\begin{document}

\thispagestyle{empty}

\renewcommand{\ehkol}{R. Zhou}
\renewcommand{\ohkol}{$r$-Matrix for the Restricted KdV Flows}

\begin{flushleft}
\footnotesize \sf
Journal of Nonlinear Mathematical Physics \qquad 1998, V.5, N~2,
\pageref{zhou-fp}--\pageref{zhou-lp}. \hfill {\sc Article}
\end{flushleft}

\vspace{-5mm}

\renewcommand{\footnoterule}{}
{\renewcommand{\thefootnote}{}
 \footnote{\prava}}

\name{{\bfseries \itshape r}-Matrix for the Restricted KdV Flows \\ with
the Neumann Constraints}\label{zhou-fp}

\Author{Ruguang ZHOU}

\Adress{Department  of Mathematics, Xuzhou Normal University,
Xuzhou 221009, P.R. China}

\Date{Received December 15, 1996; Accepted November 25, 1997}

\begin{abstract}
\noindent
    Under the Neumann constraints, each equation of the KdV hierarchy
is decomposed into two f\/inite dimensional systems, including the well-known
Neumann model. Like in  the case of the Bargmann constraint, the explicit  Lax
representations are deduced from the adjoint representation of the auxiliary
spectral problem. It is shown that the Lax operator satisf\/ies the $r$-matrix
relation in the Dirac bracket. Thus, the integrabilities of these resulting
systems with the Neumann constraints are obtained.
 \end{abstract}

\section{Introduction}
  It is well-known that  quite a few f\/inite dimensional integrable
systems (FDISs) can be obtained from soliton equation by using the
nonlinearization method of Lax pairs \cite{zhou:Cao,zhou:CG}. Following \cite{zhou:Rau}, we
call these   FDISs  the restricted f\/lows, which can be divided into two
kinds, one is the Bargmann system, which is a Hamiltonian system; and another
is the Neumann system with some constraints \cite{zhou:CG}.
Many approaches  have been developed to study the restricted f\/lows.
Among them is the $r$-matrix method.
$r$-Matrix formula contains almost necessary information of the FDISs.
In this paper, we develop  the $r$-matrix method applicable for
Neumann systems. We take restricted KdV f\/lows as an example. We f\/irst review
the results of the restricted KdV f\/lows \cite{zhou:Cao,zhou:Rau,zhou:Zeng}.
 A hierarchy of FDISs with Neumann constraints are obtained.
 We  focus  our attention on the FDISs related with  KdV equation, including
  the well-known Neumann model \cite{zhou:Neu},
which describes the motion of a particle on  an
 $(N-1)$-sphere submitted to harmonic forces.  We obtain their Lax
 representations from the adjoint representation of the  auxiliary
 linear problem  of KdV hierarchy.
All  the Lax matrices are $2\times 2$ matrices (not $N\times N$ !).
Then  in Section~3, we show  that the Lax operator
$L(\lambda)$ satisf\/ies the $r$-matrix
 relation in the  Dirac bracket.  Finally, by virtue of the general theory of
 $r$-matrix \cite{zhou:BV}, we obtain the Uhlenbeck motion integrals
 \cite{zhou:Uhl} and  prove the
 involution property in the constrained space (sphere bundle).

\section{The restricted KdV f\/lows and their Lax representations}

It is well-known that the KdV hierarchy associates the following
Schr\"odinger  spectral  problem
 \begin{equation}
  \left(\begin{array}{c} \psi_{1}\\ \psi_{2}   \end{array} \right)_x
  =U(u,\lambda)
  \left( \begin{array}{c} \psi_{1}\\ \psi_{2}   \end{array} \right),\qquad U(\lambda, u)
  =  \left( \begin{array}{cc}{0}&{1}\\
  {-\lambda-u}&{0} \end{array}\right). \label{zhou:sch}\end{equation}
 To deduce the KdV hierarchy, following \cite{zhou:Tu}, we f\/irst consider the
 adjoint representation of~(\ref{zhou:sch})
\begin{equation} V_x=[U,V]\equiv UV-VU \end{equation}
with
\begin{equation} V=  \left( \begin{array}{cc}
       {a}&{b}\\{c}&{-a} \end{array}\right)
       =\sum_{m=0}^\infty V_m\lambda^{-m}=\sum_{m=0}^\infty
  \left( \begin{array}{cc}{a_m}&{b_m}\\{c_m}&{-a_m}\end{array}\right)\lambda^{-m}  \end{equation}
 which leads to
  \begin{equation}
  b_{i+1}={\cal L}b_i, \qquad
  {\cal L}=-\frac{1}{4}\partial^2-u+\frac{1}{2}\partial^{-1}u_x,
  \end{equation}
 \begin{equation}
 a_i=-\frac{1}{2}b_{ix}, \qquad
 c_{ix}=-2a_{i+1}-2a_iu.
 \end{equation}
The f\/irst few terms are
 \begin{equation} a_0=b_0=0, \quad
 c_0=-1, \quad a_0=0,\quad b_1=1,\quad c_1=-\frac{1}{2}u,
 \end{equation}
 \begin{equation}
 a_2=\frac{1}{4}u_x,\quad b_2=-\frac{1}{2}u,
 \quad c_2=\frac{1}{8}\left(u_{xx}+u^2\right),
\quad   b_3=\frac{1}{8}\left(u_{xx}+3u^2\right),
 \end{equation}
where $\ds \partial=\frac{\partial}{\partial  x}$,
$\partial\partial^{-1}=\partial^{-1}\partial=1.$

 Then, we can take the following auxiliary spectral problem
\begin{equation}
 \left( \begin{array}{c} \psi_{1}\\ \psi_{2}   \end{array} \right)
_{t_n}=V^{(n)}(\lambda,u)
\left( \begin{array}{c} \psi_{1}\\ \psi_{2}   \end{array} \right),
\end{equation}
where
\begin{equation} V^{(n)}(\lambda,u)=\sum_{m=0}^{n+1}V_m\lambda^{n+1-m}+
  \left( \begin{array}{cc}{0}&{0}\\{b_{n+2}}&{0}\end{array}\right).
    \label{zhou:auxisch}
\end{equation}

The compatible condition of (\ref{zhou:sch}) and
(\ref{zhou:auxisch}) yields the zero-curvature
equation
\begin{equation} U_{t_n}-V_x^{(n)}+\left[U,V^{(n)}\right]=0,
\qquad n=1,2,\ldots,
\end{equation}
which gives the KdV hierarchy
\begin{equation} u_{t_n}=-2b_{n+2,x}.\end{equation}
For $n=0$, $V^{(0)}=U(u,\lambda)$.

For $n=1$ we get
\begin{equation} V^{(1)}=  \left( \begin{array}{cc}
\ds {\frac{1}{4}u_x}&\ds {\lambda-\frac{1}{2}u}\\[3mm]
      \ds  {-\lambda^2-\frac{u}{2}\lambda+\frac{u_{xx}}{4}
+\frac{u^2}{2}}& \ds{-\frac{1}{4}u_x} \end{array}\right) \label{zhou:KDVT}
\end{equation}
and the KdV equation (set $t_1=t$)
\begin{equation}
u_t=-\frac{1}{4}(u_{xxx}+6uu_x)\label{zhou:KdV}.
\end{equation}

 Now we turn to the nonlinearization of Lax pair for the KdV hierarchy.
Following \cite{zhou:Cao}, we take $N$ distinct $\lambda_j's$ and consider
the following system
\begin{equation}\psi_{1j,x}=\psi_{2j},\qquad
\psi_{2j,x}=(-\lambda_j-u)\psi_{1j}, \quad j=1,\ldots,N.
\label{zhou:Nsch}\end{equation}

 It is easy to show that (up to a constant factor)
 \begin{equation} \frac{\delta \lambda_j}{\delta u}=\psi^2_{1j},\qquad
 {\cal L}\psi^2_{1j}=\lambda_j\psi_{1j}.\label{zhou:NG}\end{equation}

 As usual \cite{zhou:Cao}, we can consider the constraints
 \begin{equation}
 b_k= \sum_{j=1}^N\frac{\delta \lambda_j}{\delta u},\qquad k=1,2,\ldots.
 \end{equation}
 In this paper we are interested in the Neumann constraint
 \begin{equation} b_1=\sum_{j=1}^N\frac{\delta \lambda_j}{\delta u},
  \end{equation}
 i.e.,
 \begin{equation}
  \langle q,q\rangle =1.
  \label{zhou:NNK}\end{equation}
 Here and hereafter, we use the notation: $q_j=\psi_{1j}$, $p_j=\psi_{2j}$,
  $q=(q_1,\ldots,q_N)^T$, $p=(p_1,\ldots,p_N)^T$ and
  $\langle \cdot,\cdot\rangle$ denotes the stand inner product in space
 ${\mathbb R}^N$.

 From (\ref{zhou:Nsch}) and (\ref{zhou:NNK}) we get
 \begin{equation}
 u=\langle p,p\rangle -\langle\Lambda q,q\rangle.
  \label{zhou:NNC}\end{equation}

   Under this constraint, (\ref{zhou:Nsch}) is nonlinearized
into the following system
 \begin{equation}
 \left\{\ba{l}
 q_x=p,\\
 p_x=-\Lambda q-(\langle p,p\rangle -\langle \Lambda q,q\rangle )q,\\
\langle q,q\rangle =1, \quad  \langle q,p\rangle =0.
 \ea
 \right.\label{zhou:KNEU}\end{equation}

When $N=3$, we have the case studied by Neumann \cite{zhou:Neu}. For
convenience, we still call~(\ref{zhou:KNEU}) the Neumann model.

Making use of the relation
\[
{\cal L}\frac{\delta  \lambda_j}{\delta u}
=\lambda_j\frac{\delta \lambda_j}{\delta u}
\]
we have
\begin{equation}
{\tilde b}_i=\langle \Lambda^{i-1}q,q\rangle,\quad
 {\tilde a}_i=-\langle \Lambda ^{i-1}q,p\rangle,
\quad {\tilde c}_i=\langle \Lambda^{i-1}p,p\rangle , \quad i=2,3,\ldots.
\end{equation}
Here and hereafter, symbol tilde means the corresponding nonlinearized
quantity
and
\begin{equation}
{\tilde V}=  \left( \begin{array}{cc}
\ds {-\sum\limits_{k=1}^N\frac{p_kq_k}{\lambda - \lambda_k}} &
\ds {\sum\limits_{k=1}^N\frac{q_k^2}{\lambda - \lambda_k}} \\[4mm]
\ds{-1-\sum\limits_{k=1}^N\frac{p_k^2}{\lambda - \lambda_k}}&
\ds {\sum\limits_{k=1}^N\frac{p_kq_k}{\lambda - \lambda_k}}\end{array}\right).
\end{equation}

 Under the condition (\ref{zhou:NNC}), the following equation still holds
 \begin{equation}
 {\tilde V}_x=[{\tilde U},{\tilde V}]. \label{zhou:NL}
 \end{equation}
Conversely a direct calculation  can verify that (\ref{zhou:NL}) is just the
Lax representation of equation (\ref{zhou:KNEU}).
For simplicity of notation, we set
$L(\lambda)=-{\tilde V}$, and $M_i={\tilde V^{(i)}}.$
Therefore we have

\medskip

\noindent
{\bf Theorem 1.} {\it The equation (\ref{zhou:KNEU}) has the Lax representation
\begin{equation}
 L_x=[M_0,L]
  \end{equation}
with
\begin{equation} L(\lambda)=\left ( \begin{array}{cc}
            A(\lambda) & B(\lambda)\\
            C(\lambda) & -A(\lambda)
            \end{array}\right),
 \quad  M_0(\lambda)=\left(\begin{array}{cc}
            0  &1\\
            -\lambda+\langle \Lambda q,q\rangle -\langle p,p\rangle  &0
            \end{array}\right),
             \end{equation}
where
\begin{equation}
 A(\lambda)=\sum_{k=1}^N\frac{p_kq_k}{\lambda - \lambda_k},\quad
 B(\lambda)=-\sum_{k=1}^N\frac{q_k^2}{\lambda - \lambda_k},\quad
 C(\lambda)=1+\sum_{k=1}^N\frac{p_k^2}{\lambda - \lambda_k}.  \end{equation}
}

    In the same way, we can discuss the nonlinearizations of the auxiliary
spectral problem
 \begin{equation}
 \left( \begin{array}{c} \psi_{1j}\\ \psi_{2j}   \end{array} \right)
 _{t_n}=V^{(n)}(\lambda_j,u)
 \left( \begin{array}{c} \psi_{1j}\\ \psi_{2j}   \end{array} \right),
 \qquad j=1,\ldots,N.
 \end{equation}
under the  constraints (\ref{zhou:NNC}) and (\ref{zhou:KNEU}) and obtain a hierarchy
of f\/inite dimensional systems  with
the Neumann constraint $\langle q,q\rangle =1$. We call this the
restricted $t_n$-f\/low.
Just like in the case of the Bargmann constraint, the identity
\begin{equation}
V_{t_n}=\left[V^{(n)}, V\right]
\end{equation}
is a nonlinearization of the Lax representation of the restricted $t_n$-f\/low.
 In particular, for $n=1$ we obtain the following system
\begin{equation}
\left\{ \begin{array}{l}
  \ds
   q_t=\Lambda p-\langle \Lambda q,p\rangle q+\frac{1}{2}\langle p, p\rangle p+
   \frac{1}{2}\langle \Lambda q,q\rangle p, \\[2mm]
\ds  p_t=-\Lambda^2q-\frac{1}{2}\langle p,p\rangle \Lambda q+
\frac{1}{2}\langle \Lambda q,q\rangle \Lambda q-
\langle \Lambda p,p\rangle q\\[2mm]
\ds \qquad  +\langle \Lambda^2q,q\rangle q-
\frac{1}{2}\langle \Lambda q,q\rangle ^2q+
\frac{1}{2}\langle p,p\rangle ^2q+\langle \Lambda q,p\rangle p,\\[2mm]
       \langle q,q\rangle =1
      \end{array}\right., \label{zhou:NN2}\end{equation}
and the Lax representation

\medskip

\noindent
{\bf Theorem 2.} {\it The equation (\ref{zhou:NN2})
possesses the following Lax representation
\begin{equation}
 L_x=[M_1,L],
 \end{equation}
\begin{equation}
 M_1(\lambda)=\left(\begin{array}{cc}
            -\langle \Lambda q,p\rangle   &\ds
            \lambda-\frac{1}{2}(\langle p,p\rangle -
            \langle \Lambda q,q\rangle )\\[2mm]
            D(\lambda)  &  \langle \Lambda q,p\rangle
            \end{array}\right),
             \end{equation}
where
\[
D(\lambda)=-\lambda^2-\frac{\langle p,p\rangle -
\langle \Lambda q,q\rangle }{2}\lambda-\langle \Lambda p, p\rangle
+\langle \Lambda^2 q,q\rangle  -\frac{\langle \Lambda q,q\rangle ^2
-\langle p,p\rangle ^2}{2}.
\]
}

\noindent
{\bf Remark.} After a simple calculation, we can show that: if $(q,p)^T$
satisf\/ies the system of equation (\ref{zhou:KNEU}) and (\ref{zhou:NN2}), then
$u=\langle p,p\rangle -\langle \Lambda q,q\rangle $ solve the  KdV equation
(\ref{zhou:KdV}).
Besides from~\cite{zhou:CG}, we know it must be a f\/inite-band solution
of the KdV equation.
Thus we can obtain the f\/inite-band solution through solving the
two f\/inite dimensional systems with the Neumann constraints.

\section{$r$-Matrix for the restricted f\/low with Neumann constraints}
    We know that the standard Poisson bracket for the two smooth functions
    $f(p,q)$, $g(p,q)$
in the symplectic space ${\mathbb R}^{2N}$,
$\ds \sum\limits_{j=1}^N dp_j\wedge dq_j $  is def\/ined by
\begin{equation}
\{f,g\}=\sum_{i=1}^N\left(\frac{\partial f}{\partial q_i}
\frac{\partial g}{\partial p_i}
-\frac{\partial f}{\partial p_i}\frac{\partial g}{\partial q_i}\right).
\end{equation}

   Since we are dealing with a constrained system with the holonomos
   constraints
\begin{equation}
F\equiv\langle q,q\rangle -1=0, \qquad G\equiv\langle q,p\rangle =0,
 \label{zhou:14}\end{equation}
i.e., the constraint manifold is
 the sphere bundle $TS^{N-1}$
\begin{equation} TS^{N-1}=\left\{(q,p)\in {\mathbb R}^{2N} \Bigl| \langle q,q\rangle =1,
\ \langle q,p\rangle =0 \right\}\end{equation},
we shall use the Dirac bracket \cite{zhou:Sund}, instead of the
standard Poisson bracket.

The Dirac bracket  for the constraint (14) is \cite{zhou:Sund}
  \begin{equation}
  \{f,g\}_D=\{f,g\}+\frac{1}{2}\{f,F\}\{G,g\}-\frac{1}{2}\{f,G\}\{F,g\}.
  \end{equation}

 A direct calculation yields
\begin{equation} \{q_k,q_l\}_D=0,\quad
 \{q_k,p_l\}_D=\delta_{kl}-q_kq_l,\quad
\{p_k, p_l\}_D=-q_kp_l+p_kq_l.
\end{equation}

\noindent
{\bf Proposition 1.}
{\it The Neumann systems
(\ref{zhou:KNEU}) and (\ref{zhou:NN2})  can be written as the following
nonstandard Hamiltonian forms
\begin{equation} q_x=\{q,H_1\}_D, \qquad p_x=\{p,H_1\}_D\end{equation}
and
\begin{equation} q_x=\{q,H_2\}_D, \qquad p_x=\{p,H_2\}_D,
\end{equation}
respectively.
Here
\begin{equation} H_1=\frac{1}{2}(\langle p,p\rangle +\langle \Lambda q,q\rangle )\end{equation}
and
\be
H_2=\frac{1}{2}\langle \Lambda^2 q,q\rangle +
\frac{1}{2}\langle \Lambda p,p\rangle
+\frac{1}{4}\langle \Lambda q,q\rangle \langle p,p\rangle
-\frac{1}{8}\langle p,p\rangle ^2-\frac{1}{8}\langle \Lambda q,q\rangle ^2.
\ee
}

It is easy to show that
\be
  \{A(\lambda), A(\mu)\}_D=\{B(\lambda), B(\mu)\}_D=0,\label{zhou:D1}
 \ee
\be
  \{C(\lambda), C(\mu)\}_D=4(A(\lambda)-A(\mu))+4(A(\mu)C(\lambda)-
  A(\lambda)C(\mu)),
\ee
\be
  \{A(\lambda), B(\mu)\}_D= \frac{2}{\lambda -\mu}(B(\lambda)-B(\mu))-
  2B(\lambda)B(\mu),
  \ee
\be
  \{A(\lambda), C(\mu)\}_D=\frac{2}{\mu -\lambda}(C(\lambda)-C(\mu))
+2B(\lambda)C(\mu)-2B(\lambda),
\ee
\be
\{B(\lambda),C(\mu)\}_D=\frac{4}{\mu-\lambda}(A(\mu)-A(\lambda))
-4B(\lambda)A(\mu),\label{zhou:D3}
 \ee
where $\lambda$, $\mu$ are two arbitrary parameters.

  We make use of the familiar notation \cite{zhou:FT}:
  $L_1(\lambda)=L(\lambda)\otimes I$, $L_2(\mu)=
 I\otimes L(\mu)$, where $I$ is the unit $2\times 2$ matrix; and  set
 \begin{equation}
 \{L_1(\lambda), L_(\mu)\}_D^{jk, mn}=\{L(\lambda)^{jm}, L(\mu)^{kn}\}_D.
 \end{equation}

It follows from (\ref{zhou:D1})--(\ref{zhou:D3}) that

\medskip

\noindent
{\bf Theorem 3.}
{\it The Lax operator $L(\lambda)$ admits the following $r$-matrix representation
 \begin{equation}
 \{L_1(\lambda),L_2(\mu)\}_D=[r_{12}(\lambda,\mu), L_1(\lambda)]-[r_{21}(\lambda, \mu), L_2(\mu)]\end{equation}
with
  \begin{equation}
  r_{12}(\lambda,\mu)=\frac{2}{\mu-\lambda}P+2S+Q_{12},
  \end{equation}
  \begin{equation}
  r_{21}(\lambda,\mu)=-\frac{2}{\mu-\lambda}P+2S+Q_{21},
  \end{equation}
where
\begin{equation}
P=\left(\begin{array}{cccc} 1&0&0&0\\ 0&0&1&0\\ 0&1&0&0\\ 0&0&0&1
            \end{array}\right), \quad
S=\left(\begin{array}{cc}{0}&{0}\\{1}&{0}\end{array}\right)\otimes
\left(\begin{array}{cc}{0}&{0}\\{1}&{0}\end{array}\right),
\end{equation}
\begin{equation}
Q_{12}=\left(\begin{array}{cc}{1}&{0}\\{0}&{0}\end{array}\right)\otimes
\left(\begin{array}{cc}{B(\mu)}&{0}\\{-2A(\mu)}&{-B(\mu)}\end{array}\right)
+\left(\begin{array}{cc}{0}&{0}\\{1}&{0}\end{array}\right)\otimes
\left(\begin{array}{cc}{0}&{B(\mu)}\\{-C(\mu)}&{0}\end{array}\right),
 \end{equation}
\begin{equation} Q_{21}=\left(\begin{array}{cc}{B(\lambda)}&{0}\\
{-2A(\lambda)}&{-B(\lambda)}\end{array}\right)\otimes
\left(\begin{array}{cc}{1}&{0}\\{0}&{0}\end{array}\right)+
           \left(\begin{array}{cc}{0}&{B(\lambda)}\\
           {-C(\lambda)}&{0}\end{array}\right)\otimes
           \left(\begin{array}{cc}{0}&{0}\\{1}&{0}\end{array}\right).
\end{equation}
}

     In order to show that $r_{12}$ is an $r$-matrix, we must check
that the bracket (24)
is  skew-symmetric and satisf\/ies the Jacobi identity.
The skew-symmetry property
follows from the relation:
\[
Q_{12}=-[M, L_2(\mu)], \quad Q_{21}(\lambda)=[M,L_1(\lambda)]
\quad \mbox{and} \quad
[[M,L_2],L_1]=[[M,L_1],L_2],
\]
where
   \begin{equation} M=\left(\begin{array}{cccc} 0& 0& 0& 0\\
                                   1& 0& 0& 0\\
                                   -1& 0& 0& 0\\
                                   0&  0& 0& 0
                \end{array}\right).\end{equation}
 Through some lengthy but straightforward calculations, we can obtain
 the Jacobi identity
  \begin{equation}
\left[\bar{L}_1,\left[r^{12}, r^{13}\right]+
\left[r^{12},r^{23}\right]\right]+\left[\bar{L}_1,\left\{L_2,r^{13}
\right\}_D - \left\{\bar{L}_3,r^{12}\right\}_D\right]+ \mbox{cycle perm.}=0,
\end{equation}
where all quantities are understood to
 ${\cal G}\otimes{\cal G}\otimes{\cal G},$
${\cal G}=sl(2)\otimes C(\lambda,\lambda^{-1})$;
and as usual the superscript denotes
the space which the corresponding tensor acts on nontrivially,
\begin{equation}
\bar{L}_1=L\otimes I\otimes I,
\qquad r^{12}=r_{12}\otimes_ I,\qquad \mbox{etc.}
\end{equation}

    An immediate consequent \cite{zhou:BV} of (24) is that
 \begin{equation}
\left\{L_1^2(\lambda), L_2^2(\mu)\right\}_D
=[\bar r_{12}(\lambda, \mu), L_1(\lambda)]-
                                 [\bar r_{21} (\lambda, \mu), L_2(\mu)] \end{equation}
with
\begin{equation} \bar r_{ij}(\lambda,\mu)=\sum_{k=0}^1\sum_{l=0}^1L_1^{1-k}(\lambda)L_2^{1-l}(\mu)
r_{ij}(\lambda,\mu)L_1^k(\lambda)L_2^l(\mu). \end{equation}
 Then it follows from (31) we have
\begin{equation}
\left\{Tr L_1^2(\lambda),Tr L_2^2(\mu)\right\}_D=
\frac{1}{4}Tr\left\{L^2(\lambda),L^2(\mu)\right\}_D=0
\end{equation}
which ensures the involution property of the integrals of motion obtained
from expanding $L^2(\lambda)$ in powers of $\lambda$.

  For our Lax pair $L(\lambda)$, we have
\begin{equation}
 -\frac{1}{2}Tr L^2(\lambda)
=- A^2(\lambda) - B(\lambda) C(\lambda)
=\sum_{k=1}^N\frac{{q_k}^2}{\lambda -\lambda_k} +\sum_{k=1}^N
\frac{1}{\lambda -\lambda_k}
{\sum_{l=1}^N}'\frac{(q_kp_l-p_kq_l)^2}{\lambda_k-\lambda_l}.
\end{equation}
Here and hereafter,
$\ds {\sum_{l=1}^N}'\ldots$ denotes the index of sum, which
is not equal to $k$.
It follows that
\begin{equation}
J_k=q_k^2+{\sum_{l=1}^N}'\frac{(q_kp_l-p_kq_l)^2}{\lambda_k-\lambda_l},
\qquad k=1,2,\ldots,N\end{equation}
are $N$ involution systems, which are just the conserved quantities obtained by Moser
and Uhlenbeck \cite{zhou:Uhl}.
  Therefore, it follows from (33) that
 \begin{equation}
  \{J_k,J_l \}_D =0,\quad k,l=1,2,\ldots,N.
  \end{equation}
 As $\ds \sum\limits_{k=1}^NJ_k=\langle q,q\rangle=1$,
 we know only $N-1$ ${J_k}'s$ are
 functionally independent among $J_1, J_2, \ldots, J_N.$

Set
 \begin{equation} F_k=\sum^N_{j=1}\lambda_j^k J_j,\end{equation}
 then $ F_k$ and $J_i$ are in involution.

 Noticing
 \begin{equation} H_0=\frac{1}{2}F_1\end{equation}
 and
 \begin{equation}
 H_1=\frac{1}{2}F_2-\frac{1}{8}F_1^2
 \end{equation}
 we know that both the system (\ref{zhou:KNEU}) and
 (\ref{zhou:NN2}) is completely integrable.

\section{Conclusion}
 In this paper, we obtained the Lax representations and the
 $r$-matrix for the restricted KdV f\/lows with the Neumann constraints.
 This approach can also be applied to other f\/inite dimensional integrable systems
 with the Neumann constraints. Besides, for all the restricted KdV
 f\/lows
which were
 discussed,  we can perform the separation of variables to integrate
 them
following
 \cite{zhou:Har}; thus obtain the f\/inite-band solution of the KdV hierarchy.

\subsection*{Acknowledgement}

I am very grateful to Professors Gu Chaohao, Hu Hesheng and Cao Cewen for their
guidance and also thanks Professor  Zhou Zixiang and Professor Qiao Zhijun
for some valuable discussions.

This project is supported by The Doctoral Programme Foundation of
Institute High Education and by the Nature Science Foundation of Education
Committee of Jiangsu province.

\label{zhou-lp}
\end{document}